\documentclass[aps,prr,twocolumn,amsmath,amssymb,superscriptaddress,floatfix,superscriptaddress]{revtex4-2}

\usepackage{graphicx}
\usepackage{multirow}
\usepackage[dvipsnames]{xcolor}
\usepackage{bm}
\usepackage{amsfonts,amssymb,amsmath}
\usepackage{graphicx,dcolumn,bm,color,braket,slashed}
\usepackage{times} 
\usepackage{comment, ulem}

\newcommand\abs[1]{\left|#1\right|}

\begin{document}

\title{Sub-symmetry Protected Topology in Topological Insulators and Superconductors}

\author{Myungjun Kang}
\affiliation{Department of Physics, Hanyang University, Seoul 04763, Korea}

\author{Mingyu Lee}
\affiliation{Department of Physics, Hanyang University, Seoul 04763, Korea}

\author{Sangmo Cheon}
\email{sangmocheon@hanyang.ac.kr}
\affiliation{Department of Physics, Hanyang University, Seoul 04763, Korea}
\affiliation{Research Institute for Natural Science, Hanyang University, Seoul 04763, Korea}

\begin{abstract}
Exploration of topology protected by a certain symmetry is central in condensed matter physics.
A recent idea of sub-symmetry-protected (SSP) topology---
remains of a broken symmetry can still protect specific topological boundary states---has been developed and demonstrated in an optical system [Nat. Phys. \textbf{19}, 992–998 (2023)].
Here, we extend this idea further by applying sub-symmetry-protecting perturbation (SSPP) to one-dimensional topological insulating and superconducting systems using the Su-Schrieffer-Hegger (SSH) and Kitaev models.
Using the tight-binding and low-energy effective theory, we show that the SSP boundary states retain topological properties while the SSPP results in the asymmetry of boundary states.
For the SSH model, an SSP zero-energy edge
state localized on one edge possesses quantized polarization.
In contrast, the other edge state is perturbed to have non-zero
energy, and its polarization is not quantized.
For topological superconductors, zero-energy SSP Majorana boundary states for spinful Kitaev models emerge on only one edge, contrary to the conventional belief that Majorana fermions emerge at opposite edges. 
Our findings can be used as a platform to expand our understanding of topological materials as they broaden our understanding of the symmetry in a topological system and a method to engineer Majorana fermions.
\end{abstract}

\maketitle

\newpage
\section{Introduction}
The topology-protecting symmetry (TPS) of electronic and superconducting systems protects the topological phases and their corresponding topological boundary states~\cite{hasan2010,chiu2016classification}.
The topological nature of such boundary states follows the bulk-boundary correspondence.
When the bulk possesses a nontrivial topological quantity, the topological boundary state arises in a finite system, which is robust to a perturbation that does not break the TPS.
Examples of such bulk properties are the polarization of the Wannier centers~\cite{resta1994modern,rabe2007physics}, the Berry phase~\cite{qi2008topological}, and the $\mathbb{Z}_2$ index first suggested by Fu and Kane~\cite{fu2007topological}.

Expanding the TPS for topological systems has been attempted using various methods.
In topological crystalline insulators, crystalline point group symmetry protects topological boundary states~\cite{fu2011topological,xu2012observation,tanaka2012experimental,munoz2016topological}.
In a certain insulator with a minuscule energy gap, a topological band crossing in the bulk is protected by the quasi-symmetry, which is a symmetry only present in the low-energy limit acting as a TPS~\cite{guo2022quasi}.
From the viewpoint of the topological Fermi liquid theory, a topological zero-energy domain-wall state protected by the quantization of the Berry phase's difference has been investigated beyond the ten-fold classification~\cite{han2023topological}.
The methods and examples mentioned are centered on finding a TPS, though somewhat unconventional, as a means to prove and protect the system's topological nature.

Here, we focus on the idea of the sub-symmetry-protected (SSP) topological characteristics under a sub-symmetry-protected perturbation (SSPP) ~\cite{poli2017partial,wang2023sub}.
This concept is centered on the premise that the TPS can be broken up into sub-symmetries, selectively protecting topological boundary states denoted as SSP boundary states.
Thus, SSP boundary states can be localized on a single boundary of a finite system while the conventional topological boundary states emerge at opposite boundaries.
Such SSP boundary states have been reported in a few electronic and optical lattice systems~\cite{wang2023sub}.
Moreover, there is no in-depth report regarding the effect of the sub-symmetry on topological superconductors and their SSP Majorana boundary states because Majorana fermions are nonintuitive and are represented by the split electron.

This work demonstrates how the sub-symmetry protects SSP boundary states for one-dimensional topological insulating and superconducting systems.
First, we examine the Su-Schrieffer-Hegger (SSH) model as a representative topological insulating system.
In this model, an SSP zero-energy edge state localized on one edge possesses quantized polarization. In contrast, the other edge state is perturbed to have non-zero energy, and its polarization is not quantized.
Next, we show the effects of the SSPP  for the one-dimensional spinless and spinful Kitaev model.
For the spinless Kitaev model, there are still two Majorana edge states under SSPP.
The wavefunction of one SSP Majorana edge state occupies only one Majorana site, similar to the conventional topological edge state of a topological superconductor. In contrast, the other edge state is perturbed, and its wavefunction occupies two Majorana fermion sites.
Finally, we confirm that the spinful Kitaev chain has zero-energy SSP Majorana edge states localized on a single edge, while the opposing edge has no zero-energy states.
This contradicts conventional topological superconductors where zero-energy Majorana states emerge at opposite edges.
If such SSP Majorana states are experimentally possible for topological superconductors, this could control the number of Majorana fermions in a system, which would be paramount for various applications to quantum devices~\cite{stern2013topological,nayak2008,ozawa2019}.

\section{1D Topological Insulator}
\begin{figure}[t]
\includegraphics[width=0.48\textwidth]{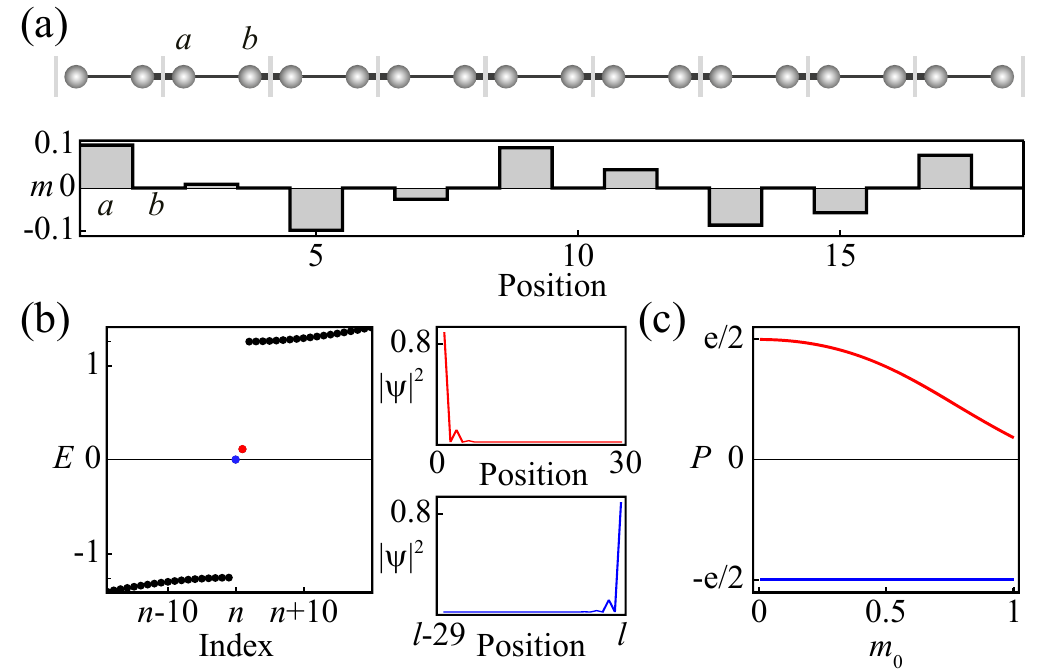}
\caption{\label{fig1:1DTI} 
\textbf{
Sub-symmetry-protected (SSP) edge state in the Su-Schrieffer-Heeger (SSH) chain.
}
\textbf{(a)} The top panel shows the schematics of the SSH chain in the topological phase.
In each unit cell indicated by vertical gray lines, there are two sublattice atoms $a$ and $b$.
The horizontal lines indicate intercell and intracell hoppings, with the thickness indicating the magnitude.
The bottom panel shows the sub-symmetry protecting perturbation (SSPP) with respect to sites.
The odd and even sites indicate the $a$ and $b$ sublattices and have non-zero and zero perturbation values, respectively.
\textbf{(b)} Energy eigenvalues near zero-energy and the wavefunctions of the in-gap states.
Blue (red) indicates the SSP (trivial) in-gap state localized on the right (left) edge.
The parameters are $t_0=\Delta=1 ,m_0=0.1$ and the length of the system is $l=200$.
\textbf{(c)} Polarization of the in-gap states with respect to the amplitude of the on-site potential, $m_0$.
The blue line indicates the polarization of the SSP in-gap state, while the red indicates the other trivial in-gap state.
}
\end{figure}

The SSH model~\cite{su1979solitons}, a well-known one-dimensional topological insulator composed of two sublattice atoms [$a$ and $b$ as indicated in Fig.~\ref{fig1:1DTI}(a)], is given by the tight-binding Hamiltonian of 
\begin{eqnarray*}
    H_{\text{SSH}} =\sum_n \left[t_0+(-1)^n\frac{\Delta}{2}\right] c^\dagger_{n+1}c_{n}+h.c.,
\end{eqnarray*}
where $c_n^\dagger$/$c_n$ indicates the creation/annihilation operator for site $n = 1,2,3, \cdots,l$, while $t_0$ and $\Delta$ indicates the hopping amplitude and dimerization, respectively.
Here, only the nearest-neighbor hopping is considered.
It is well known that topological edge states arise when the intracell hopping is smaller than the intercell hopping ($t_0>0$ and $\Delta>0$), the resulting two zero-energy edge states are localized at both ends.
In particular, the wavefunctions of the left (right) edge states reside only on the $a$ ($b$) sites.

We briefly review the TPS and possible sub-symmetries of the SSH model~\cite{wang2023sub}.
The TPS of the SSH chain is the chiral symmetry, 
\begin{eqnarray*}
    \Sigma_z H_{\text{SSH}} \Sigma_z^{-1} = - H_{\text{SSH}},
\end{eqnarray*}
where $\Sigma_z=P_a-P_b$ and $P_a$ ($P_b$) indicates the projection operator on the $a$ ($b$) sublattice.
As the sub-symmetry preserves a subset of the symmetry, the following two sub-symmetries using the projection operator on a specific sublattice can be realized in the SSH model:
\begin{eqnarray*}
    \Sigma_z H_{\text{SSH}} \Sigma_z^{-1} P_i = - H_{\text{SSH}} P_i,
\end{eqnarray*}
where $i$ can be either $a$ or $b$.
This implies that the corresponding sub-symmetry protects TPS partially only in the projected Hilbert space.
Therefore, when a perturbation that protects only one of the sub-symmetries is present, the TPS is broken, and an SSP edge state arises on a specific boundary.

We now break the TPS while protecting the sub-symmetry acting on the $b$ sites by introducing a quasi-periodic onsite potential,
\begin{eqnarray*}
    m(i) = m_0 \cos \left[2\pi\beta \left(i-1\right)\right]
\end{eqnarray*}
on the $a$ sites in the $i$th unit cell.
Here, $\beta=\sqrt{5}-1$ is the inverse golden ratio, and $m_0$ is the amplitude of the onsite potential [Fig.~\ref{fig1:1DTI}(a)].
The sub-symmetry protecting perturbation (SSPP), $m(i)$, is a highly random onsite potential, which is chosen to simulate massive disorder for $a$ sites~\cite{roy2021reentrant,roy2023critical}.

Figure~\ref{fig1:1DTI}(b) shows the numerically calculated energy eigenvalues of a system with a finite length.
As expected, one in-gap state has zero-energy, and the other in-gap state has non-zero energy.
The wavefunction distributions each indicate the in-gap state of their corresponding color and show that both in-gap states are edge states.
The SSP in-gap state (blue) is the same as the topological edge state of an SSH chain via the protection of the sub-symmetry.
On the other hand, the non-zero in-gap state (red) becomes trivial due to the perturbation and has energy approximately equal to the effective on-site potential near the left edge; $E\simeq \langle m\rangle\simeq 0.9 m_0$, which matches the first order correction given by perturbation theory [See Supplementary Material].
The SSPP affects the wavefunctions of the left edge state, as its wavefunction has non-zero values for $a$ sites when the SSPP is absent.

To see the change of the topological properties when the SSPP is introduced, we investigate the polarization of the edge states.
The modern theory of polarization indicates a connection between the polarization of a one-dimensional topological system and the system's Berry phase~\cite{resta1994modern,rabe2007physics}.
One-dimensional topological systems have a quantized fractional polarization of $\pm e/2$, where $e$ is the charge of an electron, consistent with the Berry phase difference between topological and trivial bulk phases.
The polarization of a finite chain state in respect to its unit cell is given by
\begin{eqnarray*}
    P=\frac{e}{2}\sum_n \left|\psi_{2n-1}\right|^2-\left|\psi_{2n}\right|^2,
\end{eqnarray*}
where $|\psi_n|$ is the amplitude of the wavefunction of an in-gap state at the $n$th site.
The effects of the SSPP on the polarization of the edge states can be seen in Fig.~\ref{fig1:1DTI}(c).
It is clear that the in-gap SSP state behaves similarly to a topological edge state, but the trivial in-gap state does not due to the lack of sub-symmetry to protect the state's topological characteristics.

To analytically examine the effect of the SSPP, we examine the Jakiw-Rebbi solution~\cite{jackiw1976, jackiw1981}.
The Bloch Hamiltonian for the SSH chain without any perturbation is given by
\begin{eqnarray}
\label{eq:SSH}
    H_{\text{SSH}}(k) = 2t_0 \cos k \sigma_x -\Delta \sin k \sigma_y,
\end{eqnarray}
where $\sigma_i$ is the Pauli matrix.
This Bloch Hamiltonian has time-reversal ($T=K$), particle-hole ($C=\sigma_z K$) and chiral ($\Gamma=\sigma_z$) symmetries.
These symmetries determine the class of the system according to the Altland-Zirnbauer classification table~\cite{schnyder2008classification, chiu2016classification}, which shows that the SSH chain is of the BDI class.
From the Bloch Hamiltonian, the low-energy effective Hamiltonian can be derived near the gap closing point at $k= \pm \pi/2$.
Plugging in $k=-\frac{\pi}{2} + k'$ and taking linear approximation with respect to a small $k'$, the low-energy effective Hamiltonian is given by
\begin{eqnarray*}
    H_{\text{SSH}}'(k') = 2 t_0 k' \sigma_x +\Delta \sigma_y.
\end{eqnarray*}
To describe the zero-energy Jackiw-Rebbi solution for the finite chain system, we allow $\Delta(x)$ to vary spatially at the edges such
that $\Delta(x)=\Delta_0>0$ for $0<x<l$ and $\Delta(x)=0$ elsewhere, where $l$ is the length of the system.
For this, we consider the real-space representation of the low-energy effective Hamiltonian by taking $k'=-i\partial_x$, which is 
\begin{eqnarray*}
    H_{\text{SSH}}'' = -2i t_0 \partial_x \sigma_x +\Delta(x) \sigma_y+\frac{m(x)}{2}\left(\sigma_0+\sigma_z\right),
\end{eqnarray*}
where 
$m(x)$ is the quasi-periodic on-site potential approximated in the continuum limit.
The matrix element of the SSPP shows that the term breaks the particle-hole and chiral symmetry of the system while the time-reversal symmetry remains protected.
Regardless of the form of $m(x)$, there exists a topological zero-energy Jackiw-Rebbi solution localized at the $b$ sites, which is given by
\begin{eqnarray*}
    \psi_R(x)=\mathcal{N}_R
    e^{
    -\frac{1}{2t_0}
    \int_{x}^{l} \Delta(x') dx'
    }
	\begin{pmatrix}
	0\\
	1
	\end{pmatrix}
    =\mathcal{N}_Re^{\frac{\Delta_0}{2t_0}(x-l)}
	\begin{pmatrix}
	0\\
	1
	\end{pmatrix},
\end{eqnarray*}
where $\mathcal{N}_{R}$ is a normalization factor.
As mentioned before, this solution for the SSP edge state has no value at the $a$ site.
For the other trivial in-gap state, there is no analytical solution due to the complexity of $m(x)$. 
However, if one set $m(x)=m_0$ for simplicity, an analytical solution can be obtained, which is, for energy value $E$, given by
\begin{eqnarray*}
    \psi_L(x)=\mathcal{N}_Le^{-\frac{\sqrt{E(m_0-E)+\Delta_0^2}}{2t_0}x}
	\begin{pmatrix}
	\Delta_0+\sqrt{E(m_0-E)+\Delta_0^2}\\
	i(m_0-E)
	\end{pmatrix},
\end{eqnarray*}
where $\mathcal{N}_{L}$ is a normalization factor.
Note that the solution of the trivial edge state has non-zero values in both $a$ and $b$ sites.
These analytical solutions agree well with the numerical results, which indicates that only the SPP edge state has the same topological nature as the original SSH chain.

\section{1D Topological Superconductor}
\subsection{Spinless Kitaev Chain}
We now focus on topological superconductors and their SSP Majorana fermion states.
For the simplest example, we first explore the spinless Kitaev chain, a one-dimensional topological superconductor~\cite{kitaev2001unpaired}.
The tight-binding Bogoliubov-de-Gennes (BdG) Hamiltonian of the Kitaev chain is given by
\begin{eqnarray*}
	H_{\text{BdG}}^\text{K}&=&\sum_{j} -t \left(c_{j}^\dagger c_{j+1}+c_{j+1}^\dagger c_{j}\right)-\mu c_{j}^\dagger c_{j},\\
	&&+\sum_{j} \left(\Delta_p c_{j+1} c_{j}+ h.c.\right),
\end{eqnarray*}
where $c_i^\dagger$/$c_i$ indicates the creation/annihilation operator for site $i$, while $t$ is the hopping parameter.
$\Delta_p$ is a $p$-wave pairing gap, and we choose a gauge where $\Delta_p$ is real for simplicity.
The TPS of the Kitaev chain is the chiral symmetry and the system is topological when $\abs{\mu} < \abs{2t}$.

Unlike the topological insulators, how to choose an SSPP for the system that breaks the TPS and partially protects Majorana fermions is unclear in the electron-hole representation of the BdG mean-field formalism.
However, when we take the Majorana fermion representation, which interprets one electron degree of freedom as a combination of two Majorana fermions, it is easier to determine the SSPP.

\begin{figure}[t]
\includegraphics[width=0.48\textwidth]{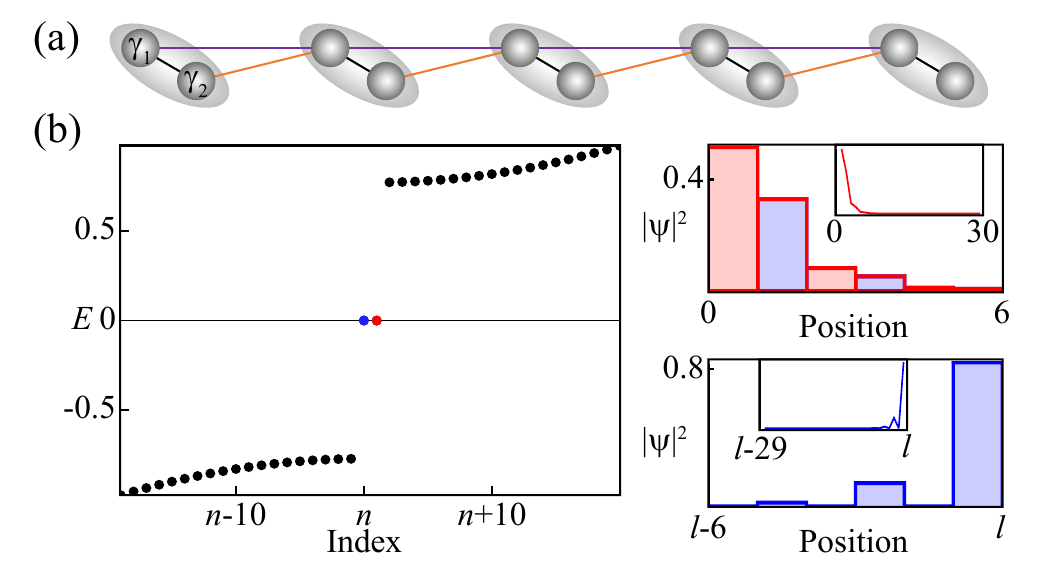}
\caption{\label{fig2:SpinlessK}
\text{\bf{SSP edge states in the spinless Kitaev chain.}}
\textbf{(a)} Schematics of the spinless Kitaev chain in the presence of SSPP using the Majorana representation.
Each physical fermion site (oval) is split into two Majorana sites (circle).
The chemical potential is given in black, the SSPP is given in purple, and both hopping and $p$-wave pairing are given in orange for simplicity.
\textbf{(b)} Energy eigenvalues and the wavefunction distributions of the zero-energy states in accordance with their color.
The wavefunction distribution is given in the Majorana representation for clarity and the order of the position is $\gamma_{1,1}, \gamma_{1,2}, \gamma_{2,1}, \cdots$, where $\gamma_{i,j}$ is the $j$th Majorana fermion at $i$th site.
The main panels (insets) are zoom-ins of the three (thirty) physical sites on the left and right for the red and blue states, respectively.
The alternating zero points for the blue state indicate it is a Majorana state, while the red state is non-Majorana, as there are non-zero distributions for neighboring Majorana sites.
The parameters are $\mu=-0.8, t=\Delta_p=1, \alpha=0.8$ and the length of the system is $l=200$.
}
\end{figure}

For this model, we use the following Majorana representation, where $i$th Majorana fermion operator at the $j$th site, $\gamma_{i,j}$, is given by $\gamma_{1,j}=\frac{1}{\sqrt{2}}\left(c_j+c_j^\dagger\right)$ or $\gamma_{2,j}=-\frac{i}{\sqrt{2}}\left(c_j-c_j^\dagger\right)$.
Then, we choose an SSPP to be a hopping term between neighboring $\gamma_{1, j}$ sites, as shown in Fig.~\ref{fig2:SpinlessK}(a), which is given as
\begin{eqnarray*}
	H_\text{SSPP}^\text{K}&=& 2\alpha \sum_{j} i \gamma_{1,j+1}\gamma_{1,j},
\end{eqnarray*}
with the strength of the SSPP, $2\alpha$.
% Note that this is similar to the additional hopping term introduced in the optical lattice version of the SSH chain~\cite{wang2023sub}.
With this, one can derive the Hamiltonian for the SSPP within the electron-hole representation, adhering to the BdG formalism, which is 
\begin{eqnarray*}
	H_\text{SSPP}^\text{K}&=&\alpha\sum_{j} \left(-ic_{j}^\dagger c_{j+1}+ic_{j+1}^\dagger c_{j}\right)\\
	&&+\alpha\sum_{j} \left(ic_{j+1} c_{j} + h.c.\right).
\end{eqnarray*}
Here, the first and second rows indicate additional hopping and pairing, respectively.
Note that the only allowed SSPP in this spinless Kitaev chain is in the form of intersite hopping between the same Majorana sites because the alternative form, an on-site potential such as $i \gamma_{i,j}^\dagger\gamma_{i,j}$, is impossible to implement because such a term is trivial mathematically.

The resulting energy eigenvalues and wavefunctions of the zero-energy states when $H_\text{SSPP}^\text{K}$ is introduced can be seen in Fig.~\ref{fig2:SpinlessK}(b).
The SSP Majorana state is given in blue, while the other is given in red.
Both energy eigenvalues remain at zero energy, which at first glance seems to indicate that both states are degenerate and, hence, SSP Majorana boundary states.
To clear this, the wavefunction distributions in the Majorana representation are examined.
Figure~\ref{fig2:SpinlessK}(b) shows the probability distribution of the wavefunction that resides in the three nearest physical sites near the left and right edges for the red and blue states, respectively.
The red wavefunction shows the distribution in both $\gamma_1$ and $\gamma_2$ Majorana sites, which implies that the red state is not a conventional Majorana edge state in the sense that it is not perfectly in neither the $\gamma_1$ nor $\gamma_2$ Majorana sites.
On the other hand, the blue wavefunction shows perfect localization on the $\gamma_2$ sites, which indicates that it is an SPP Majorana edge state similar to the conventional topological edge state of the original Kitaev chain.

The low-energy effective Hamiltonian and the resulting Jackiw-Rebbi solution are examined to check if the analytical results of the wavefunction match those of the numerical results~\cite{jackiw1976, jackiw1981}.
Taking a Fourier transformation, the tight-binding BdG Hamiltonian of the Kitaev chain with the SSPP terms will be given as $H_{\text{BdG}} = \frac{1}{2} \sum_{k} \Psi_{k}^{\dagger} H_{\text{BdG}}^{\text{K}}(k) \Psi_k$ for the basis of $\Psi_k=\left(c_k,c_{-k
}^{\dagger}\right)^T$. 
Here, the Bloch BdG Hamiltonian is 
\begin{eqnarray}
\label{eq:SLK}
	H_{\text{BdG}}^{\text{K}}(k)&=&-\left(2t\cos k +\mu\right)\tau_z+2\Delta_p \sin k \tau_y\\
	&&+2\alpha \sin k \left(\tau_0+\tau_x\right)\nonumber,
\end{eqnarray}
where $\tau_i$ is the Paluli matrix indicating the Nambu space.
From the Bloch BdG Hamiltonian, when $\alpha=0$, the system is time-reversal ($T=K$), particle-hole ($C=\tau_x K$), and chiral ($\Gamma=\tau_x$) symmetric, indicating the BDI class.
On the other hand, when $\alpha\neq0$, the time-reversal and chiral symmetry are broken, leaving the system simple particle-hole symmetric.
However, the sub-symmetry still protects the topological nature of the edge state similar to a conventional topological Majorana edge state of the BDI class.

To find the Jackiw-Rebbi solution of the Majorana states, we take the Dirac approximation near $k=0$ for the Hamiltonian in Eq.~(\ref{eq:SLK}).
Simplifying $2t+\mu$ into $\bar{\mu}$, and plugging in $-i\partial_x$ in the place of $k$, the low-energy Dirac Hamiltonian in the real space is represented as
\begin{eqnarray*}
	H_{\text{BdG}}^{{\text{K}}'}=-\bar{\mu}\tau_z-i\partial_x\left[2\Delta_p  \tau_y+2\alpha \left(\tau_0+\tau_x\right)\right].
\end{eqnarray*}
The Jackiw-Rebbi solution of the edge states localized at the left and right edges, given respectively as $\psi_L(x)$ and $\psi_R(x)$, are 
\begin{eqnarray*}
    \psi_L(x)&=&\mathcal{N}_L e^{-\frac{\bar{\mu}}{2\Delta_p}x}
	\begin{pmatrix}
	\Delta_p+i\alpha \\
	\Delta_p-i\alpha
	\end{pmatrix},\\
    \psi_R(x)&=&\mathcal{N}_R e^{\frac{\bar{\mu}}{2\Delta_p}(x-l)}
	\begin{pmatrix}
	1\\
	-1
	\end{pmatrix},
\end{eqnarray*}
where $l$ is the length of the system, $0\leq x\leq l$, and $\mathcal{N}_{L,R}$ is the normalization factor for $\psi_{L,R} (x)$.
In the Majorana representation, the wavefunctions are given by
\begin{eqnarray*}
    \psi_L(x)&=&\widetilde{\mathcal{N}}_L e^{-\frac{\bar{\mu}}{2\Delta_p}x}	\left(\Delta_p\gamma_1-\alpha\gamma_2\right),\\
    \psi_R(x)&=&\widetilde{\mathcal{N}}_R e^{\frac{\bar{\mu}}{2\Delta_p}(x-l)}\gamma_2,    
\end{eqnarray*}
where $\widetilde{\mathcal{N}}_{L,R}$ is the normalization factor.
$\psi_{R} (x)$ is a SPP Majorana fermion, independent of $\alpha$, while the Majorana fermion $\psi_{L} (x)$ is not.
These solutions are consistent with the tight-binding results as the left edge state is perturbed to have values at both Majorana sites due to the non-zero $\alpha$, and the right edge state is a Majorana state localized at the $\gamma_2$ sites, implying that it is an SPP Majorana edge state similar to the conventional topological edge state of the original Kitaev chain.

%Before closing this subsection, we would like to discuss the implications of the SSPP of the spinless Kitaev chain for general complex $\Delta_p$.
%Our results show that the relative phase of $\Delta_p$ with respect to a reference can be measured by controlling a hopping term $\alpha$ because the existence of an SSP edge is guaranteed only when the SSPP condition $\Delta_{\text{I}} = \alpha$ is satisfied.
%This is similar to the determination of the phase of the order parameter of the Goldstone mode after spontaneous symmetry breaking and the determination of the magnetization direction in ferromagnetic systems~\cite{goldstone1961field,goldstone1962broken,altland2010condensed}.

\subsection{Spinful Kitaev Chain}
\begin{figure}[t]
\includegraphics[width=0.48\textwidth]{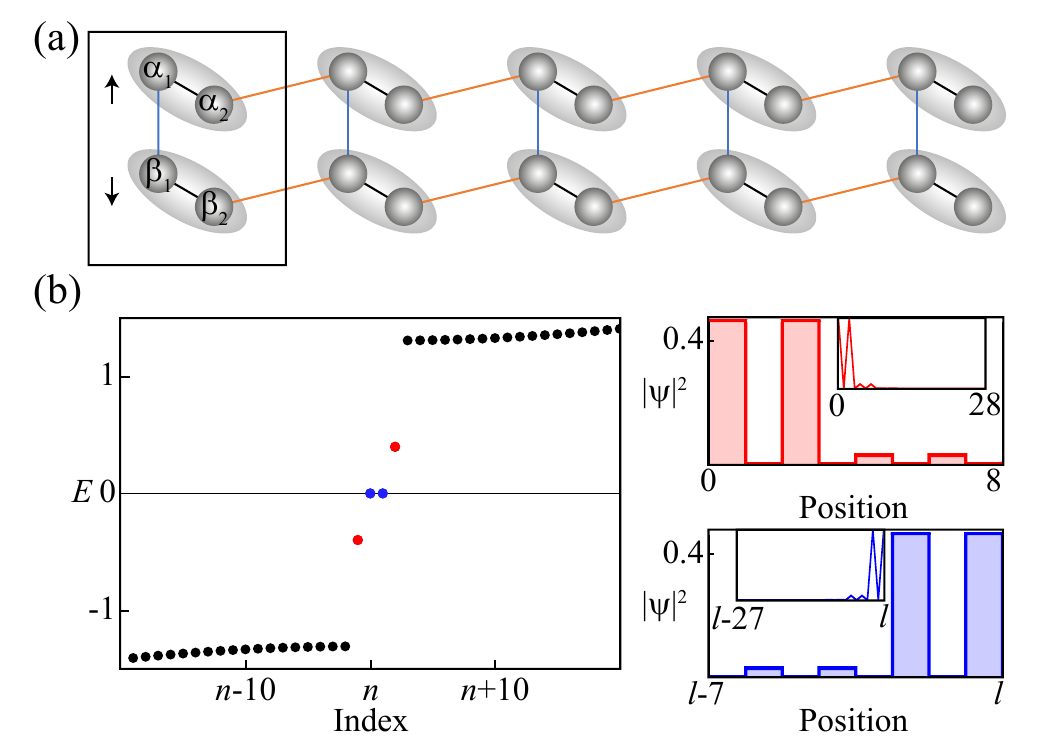}
\caption{\label{fig3:SpinfulK}
\textbf{SSP edge states in the spinful Kitaev chain.}
\textbf{(a)} Schematics of the spinful Kitaev chain in the presence of SSPP under the Majorana representation.
Each physical fermion site (oval) is split into two Majorana sites (circle).
The chemical potential is given in black, the SSPP is given in blue, and both hopping and $p$-wave pairing are given in orange for simplicity.
The square indicates the unit cell of the system with the top (bottom) row being the up (down) spin.
\textbf{(b)} Energy eigenvalues and the wavefunction distributions of the in-gap states in accordance with their color.
The wavefunction distribution is given in the Majorana representation for clarity and the order of the position is $\alpha_{1,1}, \alpha_{1,2}, \beta_{1,1}, \beta_{1,2}, \alpha_{2,1}, \cdots$, where $\gamma_{i,j}$ is the $j$th Majorana fermion at $i$th site for $\gamma=\alpha, \beta$.
The main panels (insets) are zoom-ins of the two (seven) physical sites on the left and right for the red and blue states, respectively.
The parameters are $\mu=-0.5, t=\Delta_p=1, \lambda=\Delta_s=0.2$ the total number of Majorana sites of the system is $l=400$.
}
\end{figure}

We now consider a Kitaev chain with a spin degree of freedom.
We copy the Kitaev chain given in the previous section and assign each chain with a different spin.
The SSPP is chosen as an intra-cell interaction between different Majorana fermions with different spins.
The schematics can be seen in Fig.~\ref{fig3:SpinfulK}(a), with the SSPP given in blue.
The explicit tight-binding BdG Hamiltonian is composed of the spinful Kitaev chain ($H_0^{\text{SK}}$) and the intercell SSPP ($H_\text{SSPP}^{\text{SK}}$), which is given as
\begin{eqnarray*}
    H_{\text{BdG}}^{\text{SK}}&=&H^\text{SK}_0+H_\text{SSPP}^\text{SK},\\
	H^{\text{SK}}_0 &=& \sum_{j,s} -t \left(c_{j,s}^\dagger c_{j+1,s}+c_{j+1,s}^\dagger c_{j,s}\right)-\mu c_{j,s}^\dagger c_{j,s}\\
	&&+\sum_{j,s} \left(\Delta_p^*c_{j+1,s} c_{j,s}+ h.c.\right),\\
	H_\text{SSPP}^\text{SK}&=&\sum_{j} -\lambda\left(ic_{j,\uparrow}^\dagger c_{j,\downarrow}-ic_{j,\downarrow}^\dagger c_{j,\uparrow}\right)\\
    && +\sum_{j}\left(i\Delta_s^* c_{j,\uparrow} c_{j,\downarrow}+ h.c.\right),
\end{eqnarray*}
where $c_{i,s}^\dagger$/$c_{i,s}$ indicates the creation/annihilation operator for site $i$ with spin $s$. $\lambda$ and $\Delta_s$ are the spin-flip parameter and $s$-wave pairing gap within the unit cell, respectively.
$t$ and $\Delta_p$ are the hopping amplitude and $p$-wave pairing gap, respectively.
As this model has two paring gaps, one can not make a gauge choice such that two gaps are real simultaneously.
Thus, we only choose a gauge where $\Delta_s = \abs{\Delta_s}$ is real using a global transformation of $c_i \rightarrow e^{i \theta} c_i$ and treat $\Delta_p$ as complex in general.
In the main text, however, we set $\Delta_p$ to be real for simplicity and the analytical solution for a general complex $\Delta_p$ is given in the Supplementary Material.

It should be noted that the proper SSPP only arises when $\lambda=\Delta_s$, where only $\alpha_1$ and $\beta_1$ couple, as shown in Fig.~\ref{fig3:SpinfulK}(a).
This gives the Hamiltonian for the SSPP as
\begin{eqnarray*}
	H_\text{SSPP}^\text{SK}&=& 2 \lambda \sum_{j} i \alpha_{1,j}\beta_{1,j}
\end{eqnarray*}
in the Majorana representation.
If not, a coupling exists between $\alpha_2$ and $\beta_2$, which breaks all symmetry, leaving no meaningful sub-symmetry.
This can be seen using the relation between electron-hole operators at the $j$th site ($c_j, c_j^\dagger$) and two Majorana fermion operators ($\gamma_{1,j}, \gamma_{2,j}$), which is given by 
\begin{eqnarray*}
\gamma_{1,j}=\frac{1}{\sqrt{2}}\left(c_j+c_j^\dagger\right),~~~
\gamma_{2,j}=-\frac{i}{\sqrt{2}}\left(c_j-c_j^\dagger\right),
\end{eqnarray*}
where $\gamma=\alpha, \beta$ for the up and down spin chains, respectively.
We also point out that an inter-cell SSPP is also possible; however, it will be the same SSPP as that of the spinless Kitaev chain, which we have already examined.
Therefore, we ignore the inter-cell SSPP and only enforce the intra-cell SSPP given by $H_\text{SSPP}^\text{SK}$.

The results obtained from the tight-binding Hamiltonian for the spinful Kitaev chain are quite different from those for the spinless Kitaev chain.
The energy eigenvalues of Fig.~\ref{fig3:SpinfulK}(b) present this immediately as we can see that there are not only zero but also non-zero energy in-gap states.
Figure~\ref{fig3:SpinfulK}(b) also shows the wavefunctions of such in-gap states in the order of $\alpha_{1},\alpha_{2},\beta_{1}$ and $\beta_{2}$.
The two leftmost and rightmost unit cells are chosen to see the distributions of the wavefunctions of the non-zero and zero in-gap states, respectively.
The red (blue) in-gap states indicate non-zero (zero) in-gap states localized at the left (right) edge.
The sub-symmetry of the system protects the two zero-energy Majorana states on the right edge.
The reason for the non-zero energy states of the trivial Majorana states is due to the $s$-wave pairing of the SSPP acting as Dirac mass terms, giving the states the energy value of $E=\pm 2\Delta_s$.

The low-energy effective theory is introduced to find the Jackiw-Rebbi solutions to the spinful Kitaev chain~\cite{jackiw1976, jackiw1981}.
Taking a Fourier transformation, the tight-binding BdG Hamiltonian of the spinful Kitaev chain with the SSPP terms is given by $H_{\text{BdG}} = \frac{1}{2} \sum_{k} \Psi_{k}^{\dagger} H_{\text{BdG}}^{\text{SK}}(k) \Psi_k$ for the basis of $\Psi_k=\left(c_{k,\uparrow},c_{k,\downarrow},c_{-k,\uparrow}^{\dagger},c_{-k,\downarrow}^{\dagger}\right)^T$. 
Here, the Bloch BdG Hamiltonian is given by
\begin{eqnarray}
\label{eq:SFK}
	H_{\text{BdG}}^{\text{SK}}(k)&=&-(2t \cos k +\mu)\tau_z+2\Delta_p \sin k \tau_y\\
    &&+\lambda \left(\tau_0+\tau_x\right)s_y\nonumber,
\end{eqnarray}
where the Pauli matrices $\tau_i$ and $s_j$ each indicate the Nambu space and spin degrees of freedom, respectively, and the sub-symmetry protecting condition $\lambda=\Delta_s$ is used.

The Bloch Hamiltonian of the spinful Kitaev model is time-reversal ($T=is_yK$), particle-hole ($C=i\tau_xs_y K$), and chiral ($\Gamma=\tau_x$) symmetric when the SSPP term is zero, indicating that the system is a DIII class system.
On the other hand, when SSPP exists, the time-reversal and chiral symmetry are broken, leaving the system simply particle-hole symmetric.
Therefore, the spinful Kitaev chain that is included in the DIII class transforms the D class system due to SSPP,
which seems to indicate that the SSPP induces a phase transition between the topologically non-trivial classes.
However, this transition is not the conventional topological phase transition but a sub-symmetry-induced phase transition as the topological nature of Majorana states on only one edge remains as conventional topological Majorana edge states of the DIII class.

The Jackiw-Rebbi solution of the Majorana states is found using the low-energy effective continuum Dirac Hamiltonian, which is obtained by taking the Dirac approximation near $k=0$ for Eq.~(\ref{eq:SFK}) with $k=-i\partial_x$. The resulting Dirac Hamiltonian is given by
\begin{eqnarray*}
	H_{\text{BdG}}^{\text{SK}}&=&-\bar{\mu}\tau_z-2i\Delta_p \partial_x \tau_y+\lambda \left(\tau_0+\tau_x\right)s_y.
\end{eqnarray*}
The eigenvalue equation for this Dirac Hamiltonian gives four Jackiw-Rebbi solutions.
For $E=-2\lambda$ and $E=2\lambda$, two wavefunctions, $\psi_L^1(x)$ and $\psi_L^2(x)$, are solutions localized at the left, respectively.
For $E=0$, two degenerate solutions, $\psi_R^1(x)$ and $\psi_R^2(x)$, are localized at the right.
Their explicit forms are given as 
\begin{eqnarray*}
    \psi_L^1(x)=\mathcal{N}_L^1e^{-\frac{\bar{\mu}}{2\Delta_p} x}
	&&\begin{pmatrix}
	1\\
	i\\
	1\\
	i\\
	\end{pmatrix},\\
    \psi_L^2(x)=\mathcal{N}_L^2e^{-\frac{\bar{\mu}}{2\Delta_p} x}
	&&\begin{pmatrix}
	1\\
	-i\\
	1\\
	-i\\
	\end{pmatrix},\\
    \psi_R^1(x)=\mathcal{N}_R^1e^{\frac{\bar{\mu}}{2\Delta_p} (x-l)}
	&&\begin{pmatrix}
	0\\
	1\\
	0\\
	-1\\
	\end{pmatrix},\\
    \psi_R^2(x)=\mathcal{N}_R^2e^{\frac{\bar{\mu}}{2\Delta_p} (x-l)}
	&&\begin{pmatrix}
	1\\
	0\\
	-1\\
	0\\
	\end{pmatrix},
\end{eqnarray*}
where $l$ is the length of the system, $0\leq x\leq l$, and $\mathcal{N}_{L,R}^{1,2}$ is the normalization factor.
In the Majorana representation, the wavefunctions are given as
\begin{eqnarray*}
    \psi_L^1(x)&=&\widetilde{\mathcal{N}}_L^1e^{-\frac{\bar{\mu}}{2\Delta_p} x}\left(\alpha_1+i\beta_1\right),\\
    \psi_L^2(x)&=&\widetilde{\mathcal{N}}_L^2e^{-\frac{\bar{\mu}}{2\Delta_p} x}\left(\alpha_1-i\beta_1\right),\\\
    \psi_R^1(x)&=&\widetilde{\mathcal{N}}_R^1e^{-\frac{\bar{\mu}}{2\Delta_p} (x-l)}\beta_2,\\
    \psi_R^2(x)&=&\widetilde{\mathcal{N}}_R^2e^{-\frac{\bar{\mu}}{2\Delta_p} (x-l)}\alpha_2,
\end{eqnarray*}
where $\widetilde{\mathcal{N}}_{L,R}^{1,2}$ is the normalization factor.
Thus, $\psi_L^1(x)$ and $\psi_L^2(x)$ are not Majorana fermions but a particle-hole pair with non-zero energies $E=\pm 2 \lambda$.
On the other hand,  $\psi_R^1(x)$ and $\psi_R^2(x)$ are zero-energy SSP Majorana fermions independent of $\lambda$.
Therefore, these analytical solutions are consistent with the tight-binding results.

\section{Discussion}
We now discuss the sub-symmetry protecting condition $\lambda=\Delta_s$ for the spinful Kitaev chain in Sec.~III.B.
The Bardeen-Cooper-Schrieffer (BCS) theory of superconductors indicates that the pairing gap of the superconductor is temperature-dependent~\cite{bardeen1957theory}.
Therefore, for a given $\lambda$, $\Delta_s$ can be tuned to satisfy the SSP condition $\lambda=\Delta_s$ via temperature control.
Therefore, we derive and solve the gap equation of the spinful Kitaev chain in the presence of SSPP terms and find the critical temperature $T_\text{SPP}$ at which the sub-symmetry protection condition can be realized.
However, due to the complexity of the multi-gap superconductor, we investigate the SSP condition in a restricted situation.

We consider a system where the critical temperature for the $s$-wave pairing $T_c^\text{s-wave}$ is smaller than the critical temperature for the $p$-wave pairing $T_c^\text{p-wave}$ because the $s$-wave pairing is a perturbation term for sub-symmetry and assume that the $p$-wave pairing potential is a constant and follows the BCS relation of $2\Delta_p(T=0)=3.56 k_B T_c^\text{p-wave}$.
The gap equation for the $s$-wave pairing is given by
\begin{eqnarray}
\label{eq:sw}
	1&=& \sum_k \frac{\abs{g_\text{eff}}^2}{4\varepsilon_k}\left(\tanh\frac{\lambda+\varepsilon_k}{2k_BT}+\tanh\frac{-\lambda+\varepsilon_k}{2k_BT}\right),
\end{eqnarray}
where $\varepsilon_k=\sqrt{\left(\mu+2t\cos k\right)^2+\Delta_s^2+4\Delta_p^2\sin^2 k}$ is used for simplicity~\cite{Supple}.
It is also clear that the value of the temperature-dependent $s$-wave pairing gap depends on the value of the spin-spin interaction $\lambda$.

\begin{figure}[t]
\includegraphics[width=0.48\textwidth]{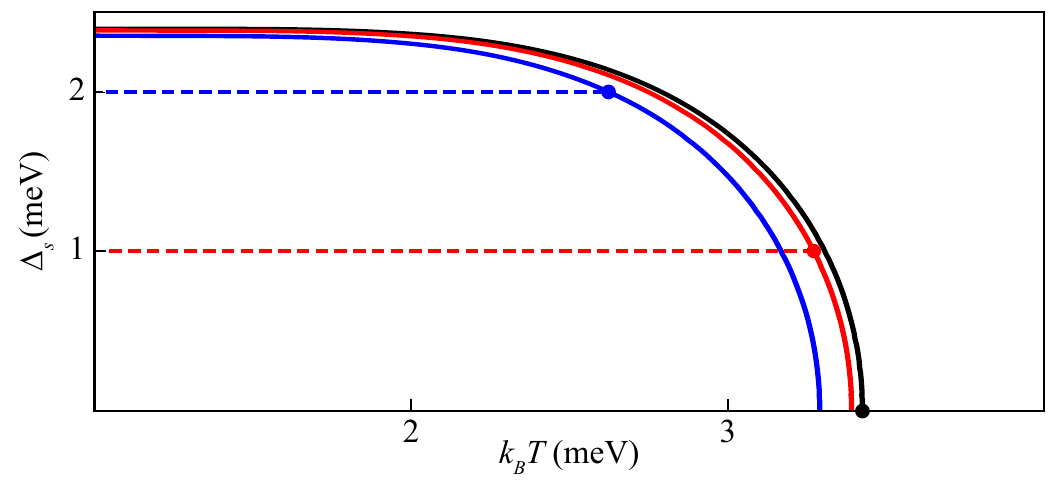}
\caption{\label{fig4:GapEq}
\textbf{Numerically calculated gap function of the $s$-wave pairing gap with respect to temperature for various $\boldsymbol{\lambda}$.}
The gap functions are plotted for $\lambda=0, 1$, and $2$ meV in black, red, and blue, respectively.
The dotted lines indicate the value of $\lambda= 1, 2$ meV.
The black, red, and blue points indicate the critical temperatures ($T_\text{SPP}$) satisfying $\Delta_s = \lambda$.
The other parameters are $t=100$ meV, $\mu=25$ meV, $\Delta_p=8$ meV, $\abs{g_\text{eff}}^2 = 82.81$ meV and $\hbar\omega_D=0.5$ meV.
}
\end{figure}

Figure~\ref{fig4:GapEq} shows the numerical results of the gap equation for various values of $\lambda$.
The plot of $\Delta_s(T)$ for various $\lambda$ is given in solid lines, and the $\lambda=\text{constant}$ graph is given in vertical dotted lines.
Therefore, when the solid and dotted lines of the same color intersect, the SSP condition is realized.
Moreover, $T_\text{SPP}$, as well as $T_c^\text{s-wave}$, decrease as $\lambda$ increases.

An exact solution of the gap equation requires solving the multi-gap equation for the $s$-and $p$-wave pairing gaps, and it is a possible candidate for future works.
However, we expect the main results will not differ greatly from our analysis.
Our result will provide a unique situation 
where Majorana boundary states arise due to the protection of sub-symmetry at a specific temperature, leading to a Majorana switch that turns the Majorana states on and off according to temperature.

\section{Conclusion}
In this work, we have investigated the effects of the sub-symmetry for various topological electronic and superconducting systems.
Even though the topology of the system is broken due to the SSPP, the SSP boundary states still emerge for specific boundaries due to the sub-symmetry.
For the SSH model, the SSP boundary states show topological characteristics possessing the quantized polarization.
The SSPP for topological superconducting systems results in asymmetric edge states for the spinless Kitaev chain. In contrast, the spinful Kitaev chain has zero-energy Majorana edge states on a single edge.

For future works, it could be possible to explore the SSPP and the corresponding SSP states beyond the gapped insulating and superconducting systems. For example, the change of the Fermi arc in a topological semi-metallic system under SSPP will be interesting.
Another possible direction is to find a more realistic model that can realize an SSPP and Majorana states in a real superconducting system.
For our spinful Kitaev system, the proximity effect-induced topological superconductor with both $s$- and $p$-wave superconductivity~\cite{fu2008superconducting,nakosai2013two,stanescu2010proximity,guan2016superconducting,chang2016topological} can be considered.
If this phenomenon were to be realized, the applications to quantum information~\cite{nayak2008, stern2013topological} and computation~\cite{lian2018topological,marra2022majorana,beenakker2020search}, which attempt to use Majorana fermions as their main information carrier, would be vast.

\section*{Acknowledgments}
This research was also supported by Quantum Simulator Development Project for Materials Innovation through the National Research Foundation of Korea(NRF) funded by the Korean government (Ministry of Science and ICT(MSIT))(No. NRF-2023M3K5A1094813).

\bibliographystyle{apsrev4-2}
% \bibliography{Ref}

\begin{thebibliography}{36}%
\makeatletter
\providecommand \@ifxundefined [1]{%
 \@ifx{#1\undefined}
}%
\providecommand \@ifnum [1]{%
 \ifnum #1\expandafter \@firstoftwo
 \else \expandafter \@secondoftwo
 \fi
}%
\providecommand \@ifx [1]{%
 \ifx #1\expandafter \@firstoftwo
 \else \expandafter \@secondoftwo
 \fi
}%
\providecommand \natexlab [1]{#1}%
\providecommand \enquote  [1]{``#1''}%
\providecommand \bibnamefont  [1]{#1}%
\providecommand \bibfnamefont [1]{#1}%
\providecommand \citenamefont [1]{#1}%
\providecommand \href@noop [0]{\@secondoftwo}%
\providecommand \href [0]{\begingroup \@sanitize@url \@href}%
\providecommand \@href[1]{\@@startlink{#1}\@@href}%
\providecommand \@@href[1]{\endgroup#1\@@endlink}%
\providecommand \@sanitize@url [0]{\catcode `\\12\catcode `\$12\catcode `\&12\catcode `\#12\catcode `\^12\catcode `\_12\catcode `\%12\relax}%
\providecommand \@@startlink[1]{}%
\providecommand \@@endlink[0]{}%
\providecommand \url  [0]{\begingroup\@sanitize@url \@url }%
\providecommand \@url [1]{\endgroup\@href {#1}{\urlprefix }}%
\providecommand \urlprefix  [0]{URL }%
\providecommand \Eprint [0]{\href }%
\providecommand \doibase [0]{https://doi.org/}%
\providecommand \selectlanguage [0]{\@gobble}%
\providecommand \bibinfo  [0]{\@secondoftwo}%
\providecommand \bibfield  [0]{\@secondoftwo}%
\providecommand \translation [1]{[#1]}%
\providecommand \BibitemOpen [0]{}%
\providecommand \bibitemStop [0]{}%
\providecommand \bibitemNoStop [0]{.\EOS\space}%
\providecommand \EOS [0]{\spacefactor3000\relax}%
\providecommand \BibitemShut  [1]{\csname bibitem#1\endcsname}%
\let\auto@bib@innerbib\@empty
%</preamble>
\bibitem [{\citenamefont {Hasan}\ and\ \citenamefont {Kane}(2010)}]{hasan2010}%
  \BibitemOpen
  \bibfield  {author} {\bibinfo {author} {\bibfnamefont {M.~Z.}\ \bibnamefont {Hasan}}\ and\ \bibinfo {author} {\bibfnamefont {C.~L.}\ \bibnamefont {Kane}},\ }\href@noop {} {\bibfield  {journal} {\bibinfo  {journal} {Rev. Mod. Phys.}\ }\textbf {\bibinfo {volume} {82}},\ \bibinfo {pages} {3045} (\bibinfo {year} {2010})}\BibitemShut {NoStop}%
\bibitem [{\citenamefont {Chiu}\ \emph {et~al.}(2016)\citenamefont {Chiu}, \citenamefont {Teo}, \citenamefont {Schnyder},\ and\ \citenamefont {Ryu}}]{chiu2016classification}%
  \BibitemOpen
  \bibfield  {author} {\bibinfo {author} {\bibfnamefont {C.-K.}\ \bibnamefont {Chiu}}, \bibinfo {author} {\bibfnamefont {J.~C.}\ \bibnamefont {Teo}}, \bibinfo {author} {\bibfnamefont {A.~P.}\ \bibnamefont {Schnyder}},\ and\ \bibinfo {author} {\bibfnamefont {S.}~\bibnamefont {Ryu}},\ }\href@noop {} {\bibfield  {journal} {\bibinfo  {journal} {Reviews of Modern Physics}\ }\textbf {\bibinfo {volume} {88}},\ \bibinfo {pages} {035005} (\bibinfo {year} {2016})}\BibitemShut {NoStop}%
\bibitem [{\citenamefont {Resta}(1994)}]{resta1994modern}%
  \BibitemOpen
  \bibfield  {author} {\bibinfo {author} {\bibfnamefont {R.}~\bibnamefont {Resta}},\ }\href@noop {} {\bibfield  {journal} {\bibinfo  {journal} {Ferroelectrics}\ }\textbf {\bibinfo {volume} {151}},\ \bibinfo {pages} {49} (\bibinfo {year} {1994})}\BibitemShut {NoStop}%
\bibitem [{\citenamefont {Rabe}\ \emph {et~al.}(2007)\citenamefont {Rabe}, \citenamefont {Ahn},\ and\ \citenamefont {Triscone}}]{rabe2007physics}%
  \BibitemOpen
  \bibfield  {author} {\bibinfo {author} {\bibfnamefont {K.~M.}\ \bibnamefont {Rabe}}, \bibinfo {author} {\bibfnamefont {C.~H.}\ \bibnamefont {Ahn}},\ and\ \bibinfo {author} {\bibfnamefont {J.-M.}\ \bibnamefont {Triscone}},\ }\href@noop {} {\emph {\bibinfo {title} {Physics of ferroelectrics: a modern perspective}}},\ Vol.\ \bibinfo {volume} {105}\ (\bibinfo  {publisher} {Springer Science \& Business Media},\ \bibinfo {year} {2007})\BibitemShut {NoStop}%
\bibitem [{\citenamefont {Qi}\ \emph {et~al.}(2008)\citenamefont {Qi}, \citenamefont {Hughes},\ and\ \citenamefont {Zhang}}]{qi2008topological}%
  \BibitemOpen
  \bibfield  {author} {\bibinfo {author} {\bibfnamefont {X.-L.}\ \bibnamefont {Qi}}, \bibinfo {author} {\bibfnamefont {T.~L.}\ \bibnamefont {Hughes}},\ and\ \bibinfo {author} {\bibfnamefont {S.-C.}\ \bibnamefont {Zhang}},\ }\href@noop {} {\bibfield  {journal} {\bibinfo  {journal} {Physical Review B}\ }\textbf {\bibinfo {volume} {78}},\ \bibinfo {pages} {195424} (\bibinfo {year} {2008})}\BibitemShut {NoStop}%
\bibitem [{\citenamefont {Fu}\ and\ \citenamefont {Kane}(2007)}]{fu2007topological}%
  \BibitemOpen
  \bibfield  {author} {\bibinfo {author} {\bibfnamefont {L.}~\bibnamefont {Fu}}\ and\ \bibinfo {author} {\bibfnamefont {C.~L.}\ \bibnamefont {Kane}},\ }\href@noop {} {\bibfield  {journal} {\bibinfo  {journal} {Physical Review B}\ }\textbf {\bibinfo {volume} {76}},\ \bibinfo {pages} {045302} (\bibinfo {year} {2007})}\BibitemShut {NoStop}%
\bibitem [{\citenamefont {Fu}(2011)}]{fu2011topological}%
  \BibitemOpen
  \bibfield  {author} {\bibinfo {author} {\bibfnamefont {L.}~\bibnamefont {Fu}},\ }\href@noop {} {\bibfield  {journal} {\bibinfo  {journal} {Physical review letters}\ }\textbf {\bibinfo {volume} {106}},\ \bibinfo {pages} {106802} (\bibinfo {year} {2011})}\BibitemShut {NoStop}%
\bibitem [{\citenamefont {Xu}\ \emph {et~al.}(2012)\citenamefont {Xu}, \citenamefont {Liu}, \citenamefont {Alidoust}, \citenamefont {Neupane}, \citenamefont {Qian}, \citenamefont {Belopolski}, \citenamefont {Denlinger}, \citenamefont {Wang}, \citenamefont {Lin}, \citenamefont {Wray} \emph {et~al.}}]{xu2012observation}%
  \BibitemOpen
  \bibfield  {author} {\bibinfo {author} {\bibfnamefont {S.-Y.}\ \bibnamefont {Xu}}, \bibinfo {author} {\bibfnamefont {C.}~\bibnamefont {Liu}}, \bibinfo {author} {\bibfnamefont {N.}~\bibnamefont {Alidoust}}, \bibinfo {author} {\bibfnamefont {M.}~\bibnamefont {Neupane}}, \bibinfo {author} {\bibfnamefont {D.}~\bibnamefont {Qian}}, \bibinfo {author} {\bibfnamefont {I.}~\bibnamefont {Belopolski}}, \bibinfo {author} {\bibfnamefont {J.}~\bibnamefont {Denlinger}}, \bibinfo {author} {\bibfnamefont {Y.}~\bibnamefont {Wang}}, \bibinfo {author} {\bibfnamefont {H.}~\bibnamefont {Lin}}, \bibinfo {author} {\bibfnamefont {L.~a.}\ \bibnamefont {Wray}}, \emph {et~al.},\ }\href@noop {} {\bibfield  {journal} {\bibinfo  {journal} {Nature communications}\ }\textbf {\bibinfo {volume} {3}},\ \bibinfo {pages} {1192} (\bibinfo {year} {2012})}\BibitemShut {NoStop}%
\bibitem [{\citenamefont {Tanaka}\ \emph {et~al.}(2012)\citenamefont {Tanaka}, \citenamefont {Ren}, \citenamefont {Sato}, \citenamefont {Nakayama}, \citenamefont {Souma}, \citenamefont {Takahashi}, \citenamefont {Segawa},\ and\ \citenamefont {Ando}}]{tanaka2012experimental}%
  \BibitemOpen
  \bibfield  {author} {\bibinfo {author} {\bibfnamefont {Y.}~\bibnamefont {Tanaka}}, \bibinfo {author} {\bibfnamefont {Z.}~\bibnamefont {Ren}}, \bibinfo {author} {\bibfnamefont {T.}~\bibnamefont {Sato}}, \bibinfo {author} {\bibfnamefont {K.}~\bibnamefont {Nakayama}}, \bibinfo {author} {\bibfnamefont {S.}~\bibnamefont {Souma}}, \bibinfo {author} {\bibfnamefont {T.}~\bibnamefont {Takahashi}}, \bibinfo {author} {\bibfnamefont {K.}~\bibnamefont {Segawa}},\ and\ \bibinfo {author} {\bibfnamefont {Y.}~\bibnamefont {Ando}},\ }\href@noop {} {\bibfield  {journal} {\bibinfo  {journal} {Nature Physics}\ }\textbf {\bibinfo {volume} {8}},\ \bibinfo {pages} {800} (\bibinfo {year} {2012})}\BibitemShut {NoStop}%
\bibitem [{\citenamefont {Munoz}\ \emph {et~al.}(2016)\citenamefont {Munoz}, \citenamefont {Vergniory}, \citenamefont {Rauch}, \citenamefont {Henk}, \citenamefont {Chulkov}, \citenamefont {Mertig}, \citenamefont {Botti}, \citenamefont {Marques},\ and\ \citenamefont {Romero}}]{munoz2016topological}%
  \BibitemOpen
  \bibfield  {author} {\bibinfo {author} {\bibfnamefont {F.}~\bibnamefont {Munoz}}, \bibinfo {author} {\bibfnamefont {M.~G.}\ \bibnamefont {Vergniory}}, \bibinfo {author} {\bibfnamefont {T.}~\bibnamefont {Rauch}}, \bibinfo {author} {\bibfnamefont {J.}~\bibnamefont {Henk}}, \bibinfo {author} {\bibfnamefont {E.~V.}\ \bibnamefont {Chulkov}}, \bibinfo {author} {\bibfnamefont {I.}~\bibnamefont {Mertig}}, \bibinfo {author} {\bibfnamefont {S.}~\bibnamefont {Botti}}, \bibinfo {author} {\bibfnamefont {M.~A.}\ \bibnamefont {Marques}},\ and\ \bibinfo {author} {\bibfnamefont {A.}~\bibnamefont {Romero}},\ }\href@noop {} {\bibfield  {journal} {\bibinfo  {journal} {Scientific Reports}\ }\textbf {\bibinfo {volume} {6}},\ \bibinfo {pages} {21790} (\bibinfo {year} {2016})}\BibitemShut {NoStop}%
\bibitem [{\citenamefont {Guo}\ \emph {et~al.}(2022)\citenamefont {Guo}, \citenamefont {Hu}, \citenamefont {Putzke}, \citenamefont {Diaz}, \citenamefont {Huang}, \citenamefont {Manna}, \citenamefont {Fan}, \citenamefont {Shekhar}, \citenamefont {Sun}, \citenamefont {Felser} \emph {et~al.}}]{guo2022quasi}%
  \BibitemOpen
  \bibfield  {author} {\bibinfo {author} {\bibfnamefont {C.}~\bibnamefont {Guo}}, \bibinfo {author} {\bibfnamefont {L.}~\bibnamefont {Hu}}, \bibinfo {author} {\bibfnamefont {C.}~\bibnamefont {Putzke}}, \bibinfo {author} {\bibfnamefont {J.}~\bibnamefont {Diaz}}, \bibinfo {author} {\bibfnamefont {X.}~\bibnamefont {Huang}}, \bibinfo {author} {\bibfnamefont {K.}~\bibnamefont {Manna}}, \bibinfo {author} {\bibfnamefont {F.-R.}\ \bibnamefont {Fan}}, \bibinfo {author} {\bibfnamefont {C.}~\bibnamefont {Shekhar}}, \bibinfo {author} {\bibfnamefont {Y.}~\bibnamefont {Sun}}, \bibinfo {author} {\bibfnamefont {C.}~\bibnamefont {Felser}}, \emph {et~al.},\ }\href@noop {} {\bibfield  {journal} {\bibinfo  {journal} {Nature physics}\ }\textbf {\bibinfo {volume} {18}},\ \bibinfo {pages} {813} (\bibinfo {year} {2022})}\BibitemShut {NoStop}%
\bibitem [{\citenamefont {Han}\ \emph {et~al.}(2023)\citenamefont {Han}, \citenamefont {Kang}, \citenamefont {Park},\ and\ \citenamefont {Cheon}}]{han2023topological}%
  \BibitemOpen
  \bibfield  {author} {\bibinfo {author} {\bibfnamefont {S.-H.}\ \bibnamefont {Han}}, \bibinfo {author} {\bibfnamefont {M.}~\bibnamefont {Kang}}, \bibinfo {author} {\bibfnamefont {M.~J.}\ \bibnamefont {Park}},\ and\ \bibinfo {author} {\bibfnamefont {S.}~\bibnamefont {Cheon}},\ }\href@noop {} {\bibfield  {journal} {\bibinfo  {journal} {arXiv preprint arXiv:2311.08771}\ } (\bibinfo {year} {2023})}\BibitemShut {NoStop}%
\bibitem [{\citenamefont {Poli}\ \emph {et~al.}(2017)\citenamefont {Poli}, \citenamefont {Schomerus}, \citenamefont {Bellec}, \citenamefont {Kuhl},\ and\ \citenamefont {Mortessagne}}]{poli2017partial}%
  \BibitemOpen
  \bibfield  {author} {\bibinfo {author} {\bibfnamefont {C.}~\bibnamefont {Poli}}, \bibinfo {author} {\bibfnamefont {H.}~\bibnamefont {Schomerus}}, \bibinfo {author} {\bibfnamefont {M.}~\bibnamefont {Bellec}}, \bibinfo {author} {\bibfnamefont {U.}~\bibnamefont {Kuhl}},\ and\ \bibinfo {author} {\bibfnamefont {F.}~\bibnamefont {Mortessagne}},\ }\href@noop {} {\bibfield  {journal} {\bibinfo  {journal} {2D Materials}\ }\textbf {\bibinfo {volume} {4}},\ \bibinfo {pages} {025008} (\bibinfo {year} {2017})}\BibitemShut {NoStop}%
\bibitem [{\citenamefont {Wang}\ \emph {et~al.}(2023)\citenamefont {Wang}, \citenamefont {Wang}, \citenamefont {Hu}, \citenamefont {Bongiovanni}, \citenamefont {Juki{\'c}}, \citenamefont {Tang}, \citenamefont {Song}, \citenamefont {Morandotti}, \citenamefont {Chen},\ and\ \citenamefont {Buljan}}]{wang2023sub}%
  \BibitemOpen
  \bibfield  {author} {\bibinfo {author} {\bibfnamefont {Z.}~\bibnamefont {Wang}}, \bibinfo {author} {\bibfnamefont {X.}~\bibnamefont {Wang}}, \bibinfo {author} {\bibfnamefont {Z.}~\bibnamefont {Hu}}, \bibinfo {author} {\bibfnamefont {D.}~\bibnamefont {Bongiovanni}}, \bibinfo {author} {\bibfnamefont {D.}~\bibnamefont {Juki{\'c}}}, \bibinfo {author} {\bibfnamefont {L.}~\bibnamefont {Tang}}, \bibinfo {author} {\bibfnamefont {D.}~\bibnamefont {Song}}, \bibinfo {author} {\bibfnamefont {R.}~\bibnamefont {Morandotti}}, \bibinfo {author} {\bibfnamefont {Z.}~\bibnamefont {Chen}},\ and\ \bibinfo {author} {\bibfnamefont {H.}~\bibnamefont {Buljan}},\ }\href@noop {} {\bibfield  {journal} {\bibinfo  {journal} {Nature physics}\ }\textbf {\bibinfo {volume} {19}},\ \bibinfo {pages} {992} (\bibinfo {year} {2023})}\BibitemShut {NoStop}%
\bibitem [{\citenamefont {Stern}\ and\ \citenamefont {Lindner}(2013)}]{stern2013topological}%
  \BibitemOpen
  \bibfield  {author} {\bibinfo {author} {\bibfnamefont {A.}~\bibnamefont {Stern}}\ and\ \bibinfo {author} {\bibfnamefont {N.~H.}\ \bibnamefont {Lindner}},\ }\href@noop {} {\bibfield  {journal} {\bibinfo  {journal} {Science}\ }\textbf {\bibinfo {volume} {339}},\ \bibinfo {pages} {1179} (\bibinfo {year} {2013})}\BibitemShut {NoStop}%
\bibitem [{\citenamefont {Nayak}\ \emph {et~al.}(2008)\citenamefont {Nayak}, \citenamefont {Simon}, \citenamefont {Stern}, \citenamefont {Freedman},\ and\ \citenamefont {Das~Sarma}}]{nayak2008}%
  \BibitemOpen
  \bibfield  {author} {\bibinfo {author} {\bibfnamefont {C.}~\bibnamefont {Nayak}}, \bibinfo {author} {\bibfnamefont {S.~H.}\ \bibnamefont {Simon}}, \bibinfo {author} {\bibfnamefont {A.}~\bibnamefont {Stern}}, \bibinfo {author} {\bibfnamefont {M.}~\bibnamefont {Freedman}},\ and\ \bibinfo {author} {\bibfnamefont {S.}~\bibnamefont {Das~Sarma}},\ }\href@noop {} {\bibfield  {journal} {\bibinfo  {journal} {Rev. Mod. Phys.}\ }\textbf {\bibinfo {volume} {80}},\ \bibinfo {pages} {1083} (\bibinfo {year} {2008})}\BibitemShut {NoStop}%
\bibitem [{\citenamefont {Ozawa}\ \emph {et~al.}(2019)\citenamefont {Ozawa}, \citenamefont {Price}, \citenamefont {Amo}, \citenamefont {Goldman}, \citenamefont {Hafezi}, \citenamefont {Lu}, \citenamefont {Rechtsman}, \citenamefont {Schuster}, \citenamefont {Simon}, \citenamefont {Zilberberg},\ and\ \citenamefont {Carusotto}}]{ozawa2019}%
  \BibitemOpen
  \bibfield  {author} {\bibinfo {author} {\bibfnamefont {T.}~\bibnamefont {Ozawa}}, \bibinfo {author} {\bibfnamefont {H.~M.}\ \bibnamefont {Price}}, \bibinfo {author} {\bibfnamefont {A.}~\bibnamefont {Amo}}, \bibinfo {author} {\bibfnamefont {N.}~\bibnamefont {Goldman}}, \bibinfo {author} {\bibfnamefont {M.}~\bibnamefont {Hafezi}}, \bibinfo {author} {\bibfnamefont {L.}~\bibnamefont {Lu}}, \bibinfo {author} {\bibfnamefont {M.~C.}\ \bibnamefont {Rechtsman}}, \bibinfo {author} {\bibfnamefont {D.}~\bibnamefont {Schuster}}, \bibinfo {author} {\bibfnamefont {J.}~\bibnamefont {Simon}}, \bibinfo {author} {\bibfnamefont {O.}~\bibnamefont {Zilberberg}},\ and\ \bibinfo {author} {\bibfnamefont {I.}~\bibnamefont {Carusotto}},\ }\href@noop {} {\bibfield  {journal} {\bibinfo  {journal} {Rev. Mod. Phys.}\ }\textbf {\bibinfo {volume} {91}},\ \bibinfo {pages} {015006} (\bibinfo {year} {2019})}\BibitemShut {NoStop}%
\bibitem [{\citenamefont {Su}\ \emph {et~al.}(1979)\citenamefont {Su}, \citenamefont {Schrieffer},\ and\ \citenamefont {Heeger}}]{su1979solitons}%
  \BibitemOpen
  \bibfield  {author} {\bibinfo {author} {\bibfnamefont {W.-P.}\ \bibnamefont {Su}}, \bibinfo {author} {\bibfnamefont {J.~R.}\ \bibnamefont {Schrieffer}},\ and\ \bibinfo {author} {\bibfnamefont {A.~J.}\ \bibnamefont {Heeger}},\ }\href@noop {} {\bibfield  {journal} {\bibinfo  {journal} {Physical review letters}\ }\textbf {\bibinfo {volume} {42}},\ \bibinfo {pages} {1698} (\bibinfo {year} {1979})}\BibitemShut {NoStop}%
\bibitem [{\citenamefont {Roy}\ \emph {et~al.}(2021)\citenamefont {Roy}, \citenamefont {Mishra}, \citenamefont {Tanatar},\ and\ \citenamefont {Basu}}]{roy2021reentrant}%
  \BibitemOpen
  \bibfield  {author} {\bibinfo {author} {\bibfnamefont {S.}~\bibnamefont {Roy}}, \bibinfo {author} {\bibfnamefont {T.}~\bibnamefont {Mishra}}, \bibinfo {author} {\bibfnamefont {B.}~\bibnamefont {Tanatar}},\ and\ \bibinfo {author} {\bibfnamefont {S.}~\bibnamefont {Basu}},\ }\href@noop {} {\bibfield  {journal} {\bibinfo  {journal} {Physical Review Letters}\ }\textbf {\bibinfo {volume} {126}},\ \bibinfo {pages} {106803} (\bibinfo {year} {2021})}\BibitemShut {NoStop}%
\bibitem [{\citenamefont {Roy}\ \emph {et~al.}(2023)\citenamefont {Roy}, \citenamefont {Nabi},\ and\ \citenamefont {Basu}}]{roy2023critical}%
  \BibitemOpen
  \bibfield  {author} {\bibinfo {author} {\bibfnamefont {S.}~\bibnamefont {Roy}}, \bibinfo {author} {\bibfnamefont {S.~N.}\ \bibnamefont {Nabi}},\ and\ \bibinfo {author} {\bibfnamefont {S.}~\bibnamefont {Basu}},\ }\href@noop {} {\bibfield  {journal} {\bibinfo  {journal} {Physical Review B}\ }\textbf {\bibinfo {volume} {107}},\ \bibinfo {pages} {014202} (\bibinfo {year} {2023})}\BibitemShut {NoStop}%
\bibitem [{\citenamefont {Jackiw}\ and\ \citenamefont {Rebbi}(1976)}]{jackiw1976}%
  \BibitemOpen
  \bibfield  {author} {\bibinfo {author} {\bibfnamefont {R.}~\bibnamefont {Jackiw}}\ and\ \bibinfo {author} {\bibfnamefont {C.}~\bibnamefont {Rebbi}},\ }\href@noop {} {\bibfield  {journal} {\bibinfo  {journal} {Phys. Rev. D}\ }\textbf {\bibinfo {volume} {13}},\ \bibinfo {pages} {3398} (\bibinfo {year} {1976})}\BibitemShut {NoStop}%
\bibitem [{\citenamefont {Jackiw}\ and\ \citenamefont {Schrieffer}(1981)}]{jackiw1981}%
  \BibitemOpen
  \bibfield  {author} {\bibinfo {author} {\bibfnamefont {R.}~\bibnamefont {Jackiw}}\ and\ \bibinfo {author} {\bibfnamefont {J.~R.}\ \bibnamefont {Schrieffer}},\ }\href@noop {} {\bibfield  {journal} {\bibinfo  {journal} {Nucl. Phys. B}\ }\textbf {\bibinfo {volume} {190}},\ \bibinfo {pages} {253} (\bibinfo {year} {1981})}\BibitemShut {NoStop}%
\bibitem [{\citenamefont {Schnyder}\ \emph {et~al.}(2008)\citenamefont {Schnyder}, \citenamefont {Ryu}, \citenamefont {Furusaki},\ and\ \citenamefont {Ludwig}}]{schnyder2008classification}%
  \BibitemOpen
  \bibfield  {author} {\bibinfo {author} {\bibfnamefont {A.~P.}\ \bibnamefont {Schnyder}}, \bibinfo {author} {\bibfnamefont {S.}~\bibnamefont {Ryu}}, \bibinfo {author} {\bibfnamefont {A.}~\bibnamefont {Furusaki}},\ and\ \bibinfo {author} {\bibfnamefont {A.~W.}\ \bibnamefont {Ludwig}},\ }\href@noop {} {\bibfield  {journal} {\bibinfo  {journal} {Physical Review B}\ }\textbf {\bibinfo {volume} {78}},\ \bibinfo {pages} {195125} (\bibinfo {year} {2008})}\BibitemShut {NoStop}%
\bibitem [{\citenamefont {Kitaev}(2001)}]{kitaev2001unpaired}%
  \BibitemOpen
  \bibfield  {author} {\bibinfo {author} {\bibfnamefont {A.~Y.}\ \bibnamefont {Kitaev}},\ }\href@noop {} {\bibfield  {journal} {\bibinfo  {journal} {Physics-uspekhi}\ }\textbf {\bibinfo {volume} {44}},\ \bibinfo {pages} {131} (\bibinfo {year} {2001})}\BibitemShut {NoStop}%
\bibitem [{\citenamefont {Bardeen}\ \emph {et~al.}(1957)\citenamefont {Bardeen}, \citenamefont {Cooper},\ and\ \citenamefont {Schrieffer}}]{bardeen1957theory}%
  \BibitemOpen
  \bibfield  {author} {\bibinfo {author} {\bibfnamefont {J.}~\bibnamefont {Bardeen}}, \bibinfo {author} {\bibfnamefont {L.~N.}\ \bibnamefont {Cooper}},\ and\ \bibinfo {author} {\bibfnamefont {J.~R.}\ \bibnamefont {Schrieffer}},\ }\href@noop {} {\bibfield  {journal} {\bibinfo  {journal} {Physical review}\ }\textbf {\bibinfo {volume} {108}},\ \bibinfo {pages} {1175} (\bibinfo {year} {1957})}\BibitemShut {NoStop}%
\bibitem [{Sup()}]{Supple}%
  \BibitemOpen
  \href@noop {} {}\bibinfo {note} {See Supplemental Material [URL] for the detailed methods, which includes Ref.~\cite{annett2004superconductivity,zhu2023topological}}\BibitemShut {NoStop}%
\bibitem [{\citenamefont {Fu}\ and\ \citenamefont {Kane}(2008)}]{fu2008superconducting}%
  \BibitemOpen
  \bibfield  {author} {\bibinfo {author} {\bibfnamefont {L.}~\bibnamefont {Fu}}\ and\ \bibinfo {author} {\bibfnamefont {C.~L.}\ \bibnamefont {Kane}},\ }\href@noop {} {\bibfield  {journal} {\bibinfo  {journal} {Physical review letters}\ }\textbf {\bibinfo {volume} {100}},\ \bibinfo {pages} {096407} (\bibinfo {year} {2008})}\BibitemShut {NoStop}%
\bibitem [{\citenamefont {Nakosai}\ \emph {et~al.}(2013)\citenamefont {Nakosai}, \citenamefont {Tanaka},\ and\ \citenamefont {Nagaosa}}]{nakosai2013two}%
  \BibitemOpen
  \bibfield  {author} {\bibinfo {author} {\bibfnamefont {S.}~\bibnamefont {Nakosai}}, \bibinfo {author} {\bibfnamefont {Y.}~\bibnamefont {Tanaka}},\ and\ \bibinfo {author} {\bibfnamefont {N.}~\bibnamefont {Nagaosa}},\ }\href@noop {} {\bibfield  {journal} {\bibinfo  {journal} {Physical Review B}\ }\textbf {\bibinfo {volume} {88}},\ \bibinfo {pages} {180503} (\bibinfo {year} {2013})}\BibitemShut {NoStop}%
\bibitem [{\citenamefont {Stanescu}\ \emph {et~al.}(2010)\citenamefont {Stanescu}, \citenamefont {Sau}, \citenamefont {Lutchyn},\ and\ \citenamefont {Sarma}}]{stanescu2010proximity}%
  \BibitemOpen
  \bibfield  {author} {\bibinfo {author} {\bibfnamefont {T.~D.}\ \bibnamefont {Stanescu}}, \bibinfo {author} {\bibfnamefont {J.~D.}\ \bibnamefont {Sau}}, \bibinfo {author} {\bibfnamefont {R.~M.}\ \bibnamefont {Lutchyn}},\ and\ \bibinfo {author} {\bibfnamefont {S.~D.}\ \bibnamefont {Sarma}},\ }\href@noop {} {\bibfield  {journal} {\bibinfo  {journal} {Physical Review B}\ }\textbf {\bibinfo {volume} {81}},\ \bibinfo {pages} {241310} (\bibinfo {year} {2010})}\BibitemShut {NoStop}%
\bibitem [{\citenamefont {Guan}\ \emph {et~al.}(2016)\citenamefont {Guan}, \citenamefont {Chen}, \citenamefont {Chu}, \citenamefont {Sankar}, \citenamefont {Chou}, \citenamefont {Jeng}, \citenamefont {Chang},\ and\ \citenamefont {Chuang}}]{guan2016superconducting}%
  \BibitemOpen
  \bibfield  {author} {\bibinfo {author} {\bibfnamefont {S.-Y.}\ \bibnamefont {Guan}}, \bibinfo {author} {\bibfnamefont {P.-J.}\ \bibnamefont {Chen}}, \bibinfo {author} {\bibfnamefont {M.-W.}\ \bibnamefont {Chu}}, \bibinfo {author} {\bibfnamefont {R.}~\bibnamefont {Sankar}}, \bibinfo {author} {\bibfnamefont {F.}~\bibnamefont {Chou}}, \bibinfo {author} {\bibfnamefont {H.-T.}\ \bibnamefont {Jeng}}, \bibinfo {author} {\bibfnamefont {C.-S.}\ \bibnamefont {Chang}},\ and\ \bibinfo {author} {\bibfnamefont {T.-M.}\ \bibnamefont {Chuang}},\ }\href@noop {} {\bibfield  {journal} {\bibinfo  {journal} {Science advances}\ }\textbf {\bibinfo {volume} {2}},\ \bibinfo {pages} {e1600894} (\bibinfo {year} {2016})}\BibitemShut {NoStop}%
\bibitem [{\citenamefont {Chang}\ \emph {et~al.}(2016)\citenamefont {Chang}, \citenamefont {Chen}, \citenamefont {Bian}, \citenamefont {Huang}, \citenamefont {Zheng}, \citenamefont {Neupert}, \citenamefont {Sankar}, \citenamefont {Xu}, \citenamefont {Belopolski}, \citenamefont {Chang} \emph {et~al.}}]{chang2016topological}%
  \BibitemOpen
  \bibfield  {author} {\bibinfo {author} {\bibfnamefont {T.-R.}\ \bibnamefont {Chang}}, \bibinfo {author} {\bibfnamefont {P.-J.}\ \bibnamefont {Chen}}, \bibinfo {author} {\bibfnamefont {G.}~\bibnamefont {Bian}}, \bibinfo {author} {\bibfnamefont {S.-M.}\ \bibnamefont {Huang}}, \bibinfo {author} {\bibfnamefont {H.}~\bibnamefont {Zheng}}, \bibinfo {author} {\bibfnamefont {T.}~\bibnamefont {Neupert}}, \bibinfo {author} {\bibfnamefont {R.}~\bibnamefont {Sankar}}, \bibinfo {author} {\bibfnamefont {S.-Y.}\ \bibnamefont {Xu}}, \bibinfo {author} {\bibfnamefont {I.}~\bibnamefont {Belopolski}}, \bibinfo {author} {\bibfnamefont {G.}~\bibnamefont {Chang}}, \emph {et~al.},\ }\href@noop {} {\bibfield  {journal} {\bibinfo  {journal} {Physical Review B}\ }\textbf {\bibinfo {volume} {93}},\ \bibinfo {pages} {245130} (\bibinfo {year} {2016})}\BibitemShut {NoStop}%
\bibitem [{\citenamefont {Lian}\ \emph {et~al.}(2018)\citenamefont {Lian}, \citenamefont {Sun}, \citenamefont {Vaezi}, \citenamefont {Qi},\ and\ \citenamefont {Zhang}}]{lian2018topological}%
  \BibitemOpen
  \bibfield  {author} {\bibinfo {author} {\bibfnamefont {B.}~\bibnamefont {Lian}}, \bibinfo {author} {\bibfnamefont {X.-Q.}\ \bibnamefont {Sun}}, \bibinfo {author} {\bibfnamefont {A.}~\bibnamefont {Vaezi}}, \bibinfo {author} {\bibfnamefont {X.-L.}\ \bibnamefont {Qi}},\ and\ \bibinfo {author} {\bibfnamefont {S.-C.}\ \bibnamefont {Zhang}},\ }\href@noop {} {\bibfield  {journal} {\bibinfo  {journal} {Proceedings of the National Academy of Sciences}\ }\textbf {\bibinfo {volume} {115}},\ \bibinfo {pages} {10938} (\bibinfo {year} {2018})}\BibitemShut {NoStop}%
\bibitem [{\citenamefont {Marra}(2022)}]{marra2022majorana}%
  \BibitemOpen
  \bibfield  {author} {\bibinfo {author} {\bibfnamefont {P.}~\bibnamefont {Marra}},\ }\href@noop {} {\bibfield  {journal} {\bibinfo  {journal} {Journal of Applied Physics}\ }\textbf {\bibinfo {volume} {132}} (\bibinfo {year} {2022})}\BibitemShut {NoStop}%
\bibitem [{\citenamefont {Beenakker}(2020)}]{beenakker2020search}%
  \BibitemOpen
  \bibfield  {author} {\bibinfo {author} {\bibfnamefont {C.}~\bibnamefont {Beenakker}},\ }\href@noop {} {\bibfield  {journal} {\bibinfo  {journal} {SciPost Physics Lecture Notes}\ ,\ \bibinfo {pages} {015}} (\bibinfo {year} {2020})}\BibitemShut {NoStop}%
\bibitem [{\citenamefont {Annett}(2004)}]{annett2004superconductivity}%
  \BibitemOpen
  \bibfield  {author} {\bibinfo {author} {\bibfnamefont {J.~F.}\ \bibnamefont {Annett}},\ }\href@noop {} {\emph {\bibinfo {title} {Superconductivity, superfluids and condensates}}},\ Vol.~\bibinfo {volume} {5}\ (\bibinfo  {publisher} {Oxford University Press},\ \bibinfo {year} {2004})\BibitemShut {NoStop}%
\bibitem [{\citenamefont {Zhu}\ \emph {et~al.}(2023)\citenamefont {Zhu}, \citenamefont {Zhuang}, \citenamefont {Wu},\ and\ \citenamefont {Yan}}]{zhu2023topological}%
  \BibitemOpen
  \bibfield  {author} {\bibinfo {author} {\bibfnamefont {D.}~\bibnamefont {Zhu}}, \bibinfo {author} {\bibfnamefont {Z.-Y.}\ \bibnamefont {Zhuang}}, \bibinfo {author} {\bibfnamefont {Z.}~\bibnamefont {Wu}},\ and\ \bibinfo {author} {\bibfnamefont {Z.}~\bibnamefont {Yan}},\ }\href@noop {} {\bibfield  {journal} {\bibinfo  {journal} {Physical Review B}\ }\textbf {\bibinfo {volume} {108}},\ \bibinfo {pages} {184505} (\bibinfo {year} {2023})}\BibitemShut {NoStop}%
\end{thebibliography}

%

\end{document}